\documentclass[manuscript]{acmart}
\usepackage{subcaption}

\AtBeginDocument{%
  }

\acmDOI{}

\acmConference[XAIxArts 2026]{Explainable AI for the Arts Workshop 2026}{July 13, 2026}{London, UK}

\begin{document}

\title{Diffusion TV: Experiencing Diffusion Models through Tangible, Embodied Interaction}


\author{Sihwa Park}
\email{shpark@yorku.ca}
\orcid{0000-0001-9620-8903}
\affiliation{
    \institution{York University}
    \city{Toronto}
    \state{ON}
    \country{Canada}
}

\begin{abstract}
  Diffusion TV is an interactive AI art installation that offers a tangible and embodied experience of diffusion models through a modified CRT TV. By physically manipulating the TV’s antenna, audiences control the clarity of AI-generated images and sounds, metaphorically enacting the denoising process that underlies diffusion-based generation. Using the tuning knob, participants switch between three channels featuring AI-generated animals from the Past (extinct species), Present (endangered species), and Future (speculative creatures), situating the interaction within a temporal and ecological narrative. Through continuous audiovisual feedback and physical interaction, Diffusion TV foregrounds the generative process over final outputs, allowing audiences to explore intermediate states as experiential material. Rather than providing explicit technical explanation, the work presents an alternative, embodied mode of explainable AI that invites exploratory engagement with and reflection on generative technologies.
\end{abstract}




\maketitle

\section{Introduction and Background}
Since the introduction of Denoising Diffusion Probabilistic Models (DDPMs) by Ho et al.~\cite{hoDenoisingDiffusionProbabilistic2020}, diffusion-based approaches have become a dominant paradigm for generative modeling across image, audio, and video. These models frame generation as an iterative transformation from noise to structured data, making the process temporally observable. In explainable AI (XAI)~\cite{barredoarrietaExplainableArtificialIntelligence2020}, intermediate outputs produced during this denoising process are generally used to analyze model behavior and interpret how structure emerges over time~\cite{parkExplainingGenerativeDiffusion2024}. Outside research contexts, however, such intermediate states are rarely exposed to broader audiences and are typically presented as pre-rendered or static visualizations, limiting opportunities for interactive or embodied exploration.

While XAI has been widely studied in functional and task-oriented domains, its application in creative contexts remains comparatively underexplored~\cite{bryan-kinnsExplainableAIArts2023}. As generative AI increasingly shapes cultural production, questions arise as to how its processes can be made legible and meaningful to non-experts beyond conventional technical explanation, which commonly relies on 2D, screen-based graphical user interfaces (GUIs)~\cite{colleyTangibleExplainableAI2022, ghajargarGraspableAIPhysical2022}. In this context, alternative approaches have emerged. The XAIxArts workshop series, for example, emphasizes artistic practice as a means of explainability, advancing alternatives to technocentric XAI within the arts~\cite{bryan-kinnsXAIxArtsManifestoExplainable2025}. Hemment et al.~\cite{hemmentExperientialAIArts2024} introduce experiential AI as an approach that leverages artistic practices and tangible experiences to mediate between opaque computational systems and human understanding. Tangible explainable AI~\cite{colleyTangibleExplainableAI2022} and graspable AI~\cite{ghajargarGraspableAIPhysical2022} demonstrate how physical interaction, material metaphors, and aesthetic qualities can support experiential understanding, allowing users to engage with AI systems not only intellectually but also perceptually and emotionally.

This paper presents Diffusion TV, an interactive AI art installation that explores alternative ways of engaging with complex AI mechanisms through embodied interaction. Using a modified cathode-ray tube (CRT) television (TV), the work enables audiences to interact with the denoising process of diffusion models by mapping antenna and knob interactions to intermediate stages of image and sound generation. In doing so, Diffusion TV reframes denoising as a performative and exploratory act, treating intermediate states as primary experiential material. Rather than prioritizing precise technical explanation, the work emphasizes intuitive, sensory engagement, inviting audiences to explore generative processes and reflect on broader themes of disappearance, preservation, and the relationship between humanity, technology, and the environment.
\section{Diffusion TV Design}

\begin{figure}[ht]
  \centering
  \begin{subfigure}[b]{0.24\textwidth}
    \centering
    \includegraphics[width=\linewidth]{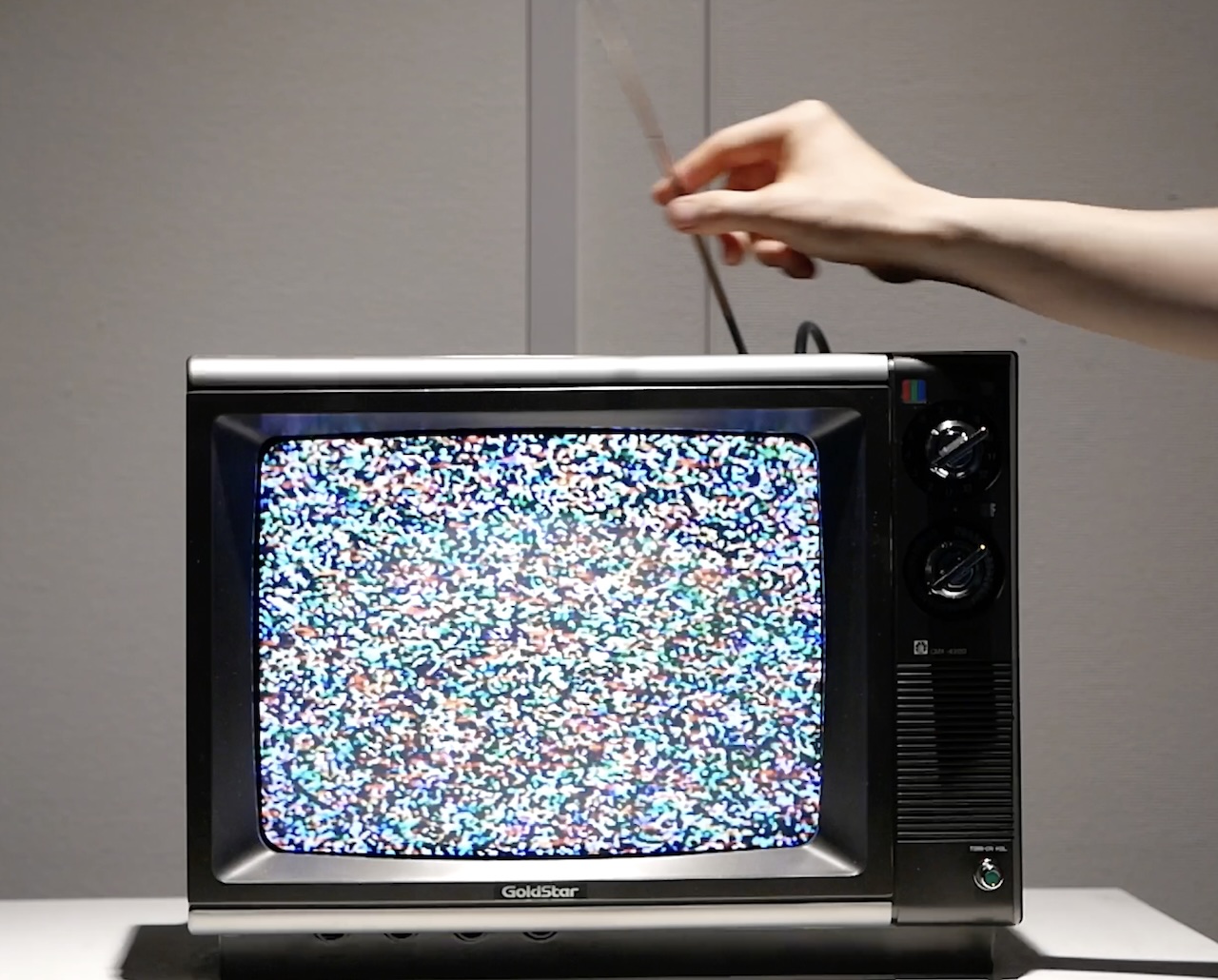}
    \caption{}
    \label{fig:antenna_interaction_a}
  \end{subfigure}\hfill
  \begin{subfigure}[b]{0.24\textwidth}
    \centering
    \includegraphics[width=\linewidth]{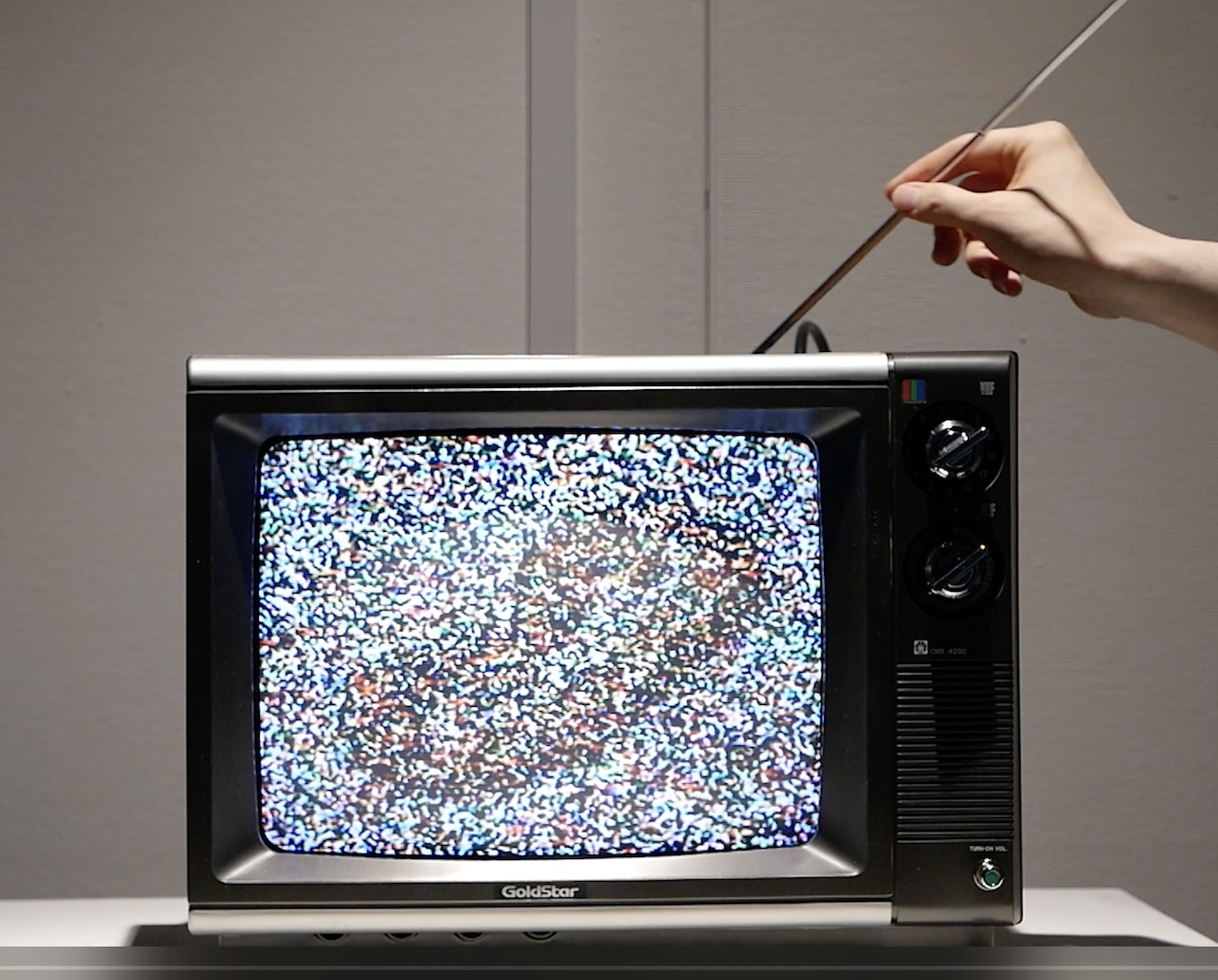}
    \caption{}
    \label{fig:antenna_interaction_b}
  \end{subfigure}\hfill
  \begin{subfigure}[b]{0.24\textwidth}
    \centering
    \includegraphics[width=\linewidth]{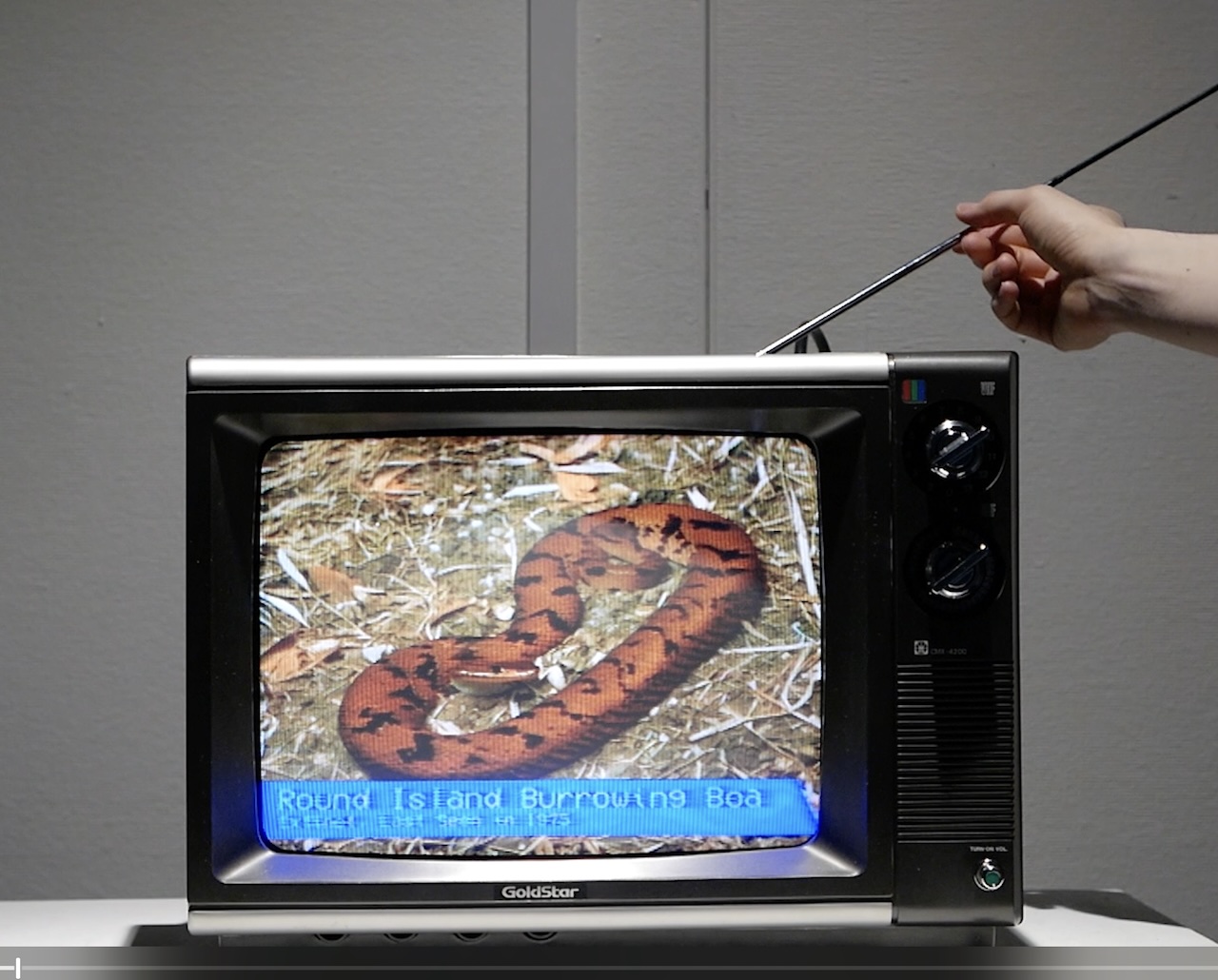}
    \caption{}
    \label{fig:antenna_interaction_c}
  \end{subfigure}\hfill
  \begin{subfigure}[b]{0.24\textwidth}
    \centering
    \includegraphics[width=\linewidth]{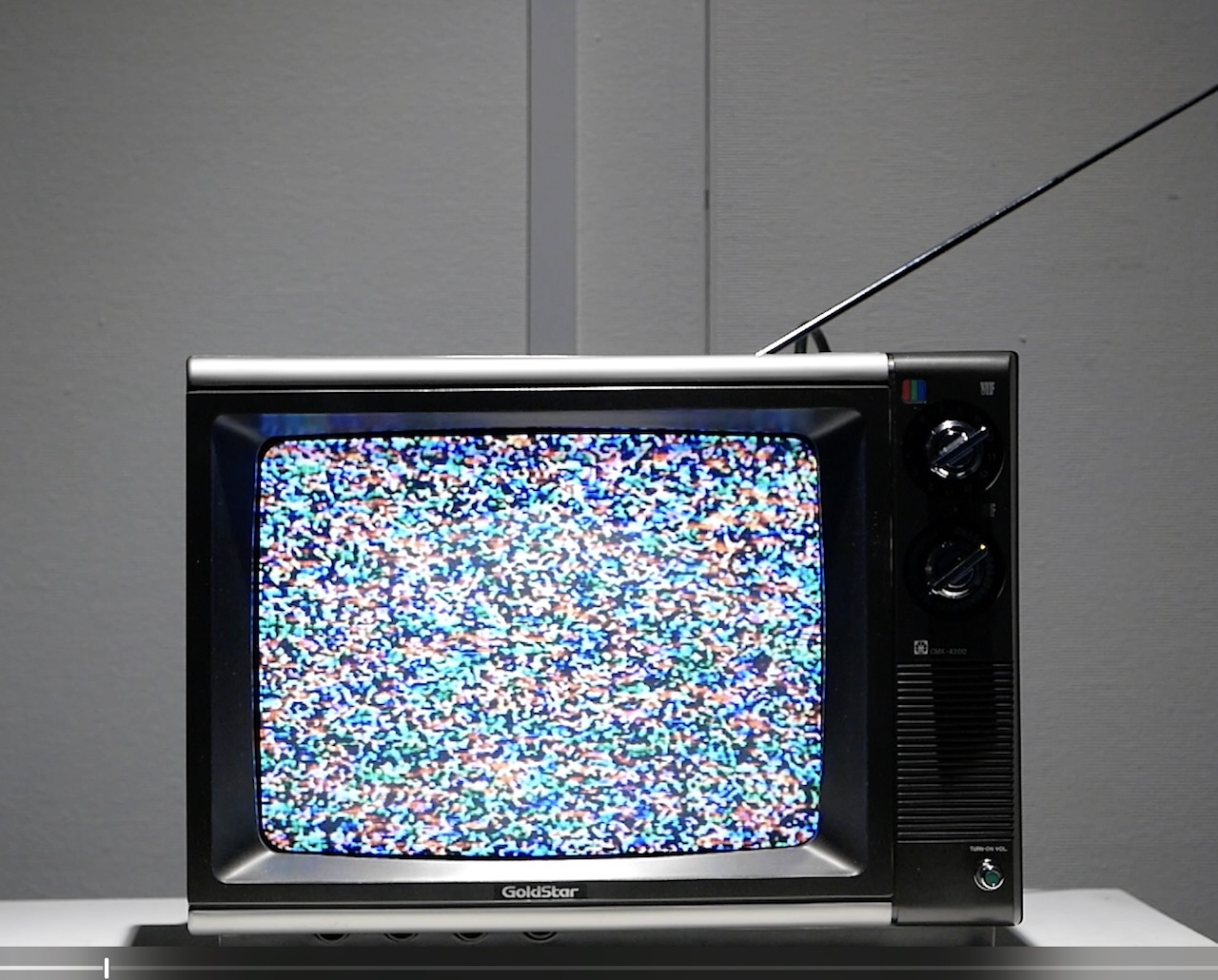}
    \caption{}
    \label{fig:antenna_interaction_d}
  \end{subfigure}
  \caption{Content display sequence with antenna interaction: (a) the noisiest image of an animal, (b) an intermediate denoising state as the antenna rotates, (c) the final denoised image with additional information, and (d) when no rotation input is detected, the noisiest image of the next animal appears after a few seconds.}
  \Description{Images showing how content updates in response to antenna rotation.}
  \label{fig:antenna_interaction}
\end{figure}

Situating the work within traditions of media archaeology and artistic reuse of outdated technologies~\cite{parikka2012media}, Diffusion TV employs the format of a nostalgic CRT TV as both interface and conceptual framework. The physical affordances of CRT interaction, such as tuning knobs and antennas, are mapped to the denoising process of diffusion models, allowing audiences to engage with generative transformation through familiar broadcast metaphors.

By rotating the antenna, viewers control the visual and auditory clarity of the AI-generated images and sounds of an animal, moving across intermediate stages from noise to structured output. Once a denoising sequence reaches clarity, the system resets after a short interval, introducing a new animal and maintaining a continuous, evolving interaction (see Figure~\ref{fig:antenna_interaction}).

The content is organized through a multichannel structure accessible via the very-high frequency (VHF) tuner knob, consisting of three channels: \textit{Past}, \textit{Present}, and \textit{Future}. The \textit{Past} channel presents extinct animals, evoking a sense of loss and remembrance. The \textit{Present} focuses on endangered animals, offering a moment of reflection on the current environmental crisis. The \textit{Future} channel features speculative creatures imagined with AI, encouraging contemplation about where humanity and technology are headed. This structure situates the generative process within temporal and ecological narratives.

At the point of clarity, brief contextual information appears in a lower-third format reminiscent of broadcast TV. Varying by channel, this information provides minimal cues, such as identifiers, temporal references, or speculative descriptors. Moreover, analogous to conventional television programming, Diffusion TV continuously regenerates the images and sounds of the animals at configurable intervals, such as daily updates. This content update ensures repeated visits yield different audiovisual experiences.

\section{Animal Data}
Several sources, including the International Union for Conservation of Nature’s Red List of Threatened Species~\cite{iucn_redlist} and the World Wildlife Fund website~\cite{wwf}, were used to identify sixteen extinct and fourteen endangered species, along with contextual information such as last-seen year, habitat, and estimated population size. Image and sound generation prompts were manually designed and iteratively refined for each species. For the \textit{Future} channel, twelve speculative species were generated using ChatGPT o3~\cite{chatgpt}, including each species’ name, emergence year, environmental trigger, short bio, and prompts for image and sound generation.

\section{Technical Overview}
\begin{figure}[ht]
  \centering
  \includegraphics[width=0.8\textwidth]{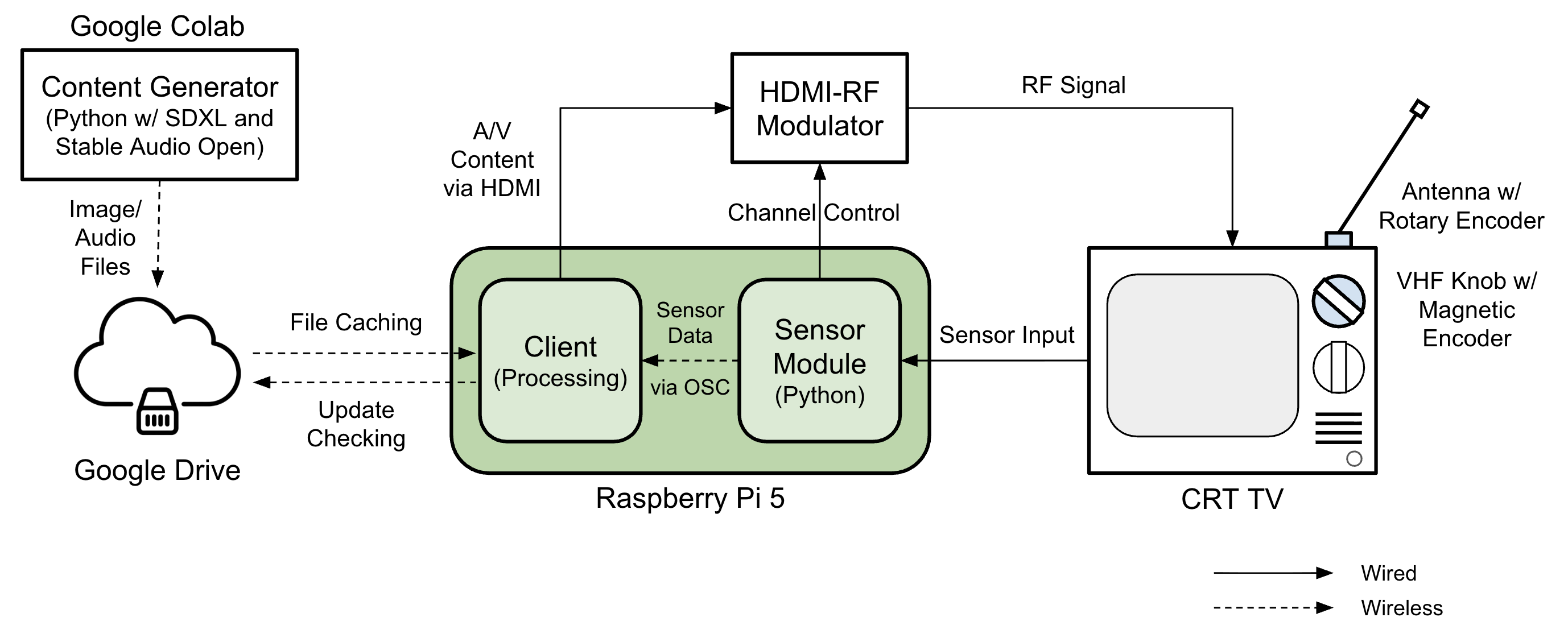}
  \caption{Schematic diagram of Diffusion TV}
  \Description{Schematic diagram illustrating components of Diffusion TV, their connections, and data flows, including a Python content generator in Google Colab, a shared Google Drive storage, a Raspberry Pi 5 running a Processing client program and Python sensor module, an HDMI-RF modulator, and a CRT TV with rotary and magnetic encoders.}
  \label{fig:diagram}
\end{figure}

As Figure~\ref{fig:diagram} illustrates, Diffusion TV consists of a modified CRT TV (GoldStar CMX-4200 13", 1987), a Raspberry Pi 5 (RPi) running a custom client program, and a sensor system for capturing user interaction. The RPi outputs the generated audiovisual content through a modified HDMI-to-radio frequency (RF) modulator for display on the TV. 

Images and sounds for all animals were pre-generated with Stable Diffusion XL~\cite{podell2024sdxl} and Stable Audio Open~\cite{Evans2025stableaudio}, using the animal dataset and custom Python code executed in Google Colab~\cite{colab}. Model inference was performed using the Diffusers~\cite{von-platen-etal-2022-diffusers} library, with custom callback functions used to extract intermediate outputs at each denoising step. These outputs were stored and made accessible to the client via Google Drive~\cite{gdrive}.

User interaction is captured through rotary and magnetic encoders integrated into the TV. A telescopic antenna is mechanically linked to a rotary encoder via a custom 3D-printed connector, mapping horizontal rotation to a continuous control signal. This signal is mapped to an index that selects corresponding image and sound pairs from different denoising stages. Channel selection is detected via a magnetic encoder attached to the VHF tuner, enabling switching between content sets associated with the \textit{Past}, \textit{Present}, and \textit{Future} channels.

The client, developed in Processing~\cite{processing}, manages content playback and interaction mapping. It dynamically updates audiovisual output in response to user input, selecting pre-generated media based on antenna position and channel state. The client also periodically checks for newly generated content and updates the dataset asynchronously without interrupting interaction.

\section{Exhibition and Discussion}
\begin{figure}[ht]
  \centering
  \includegraphics[width=0.7\textwidth]{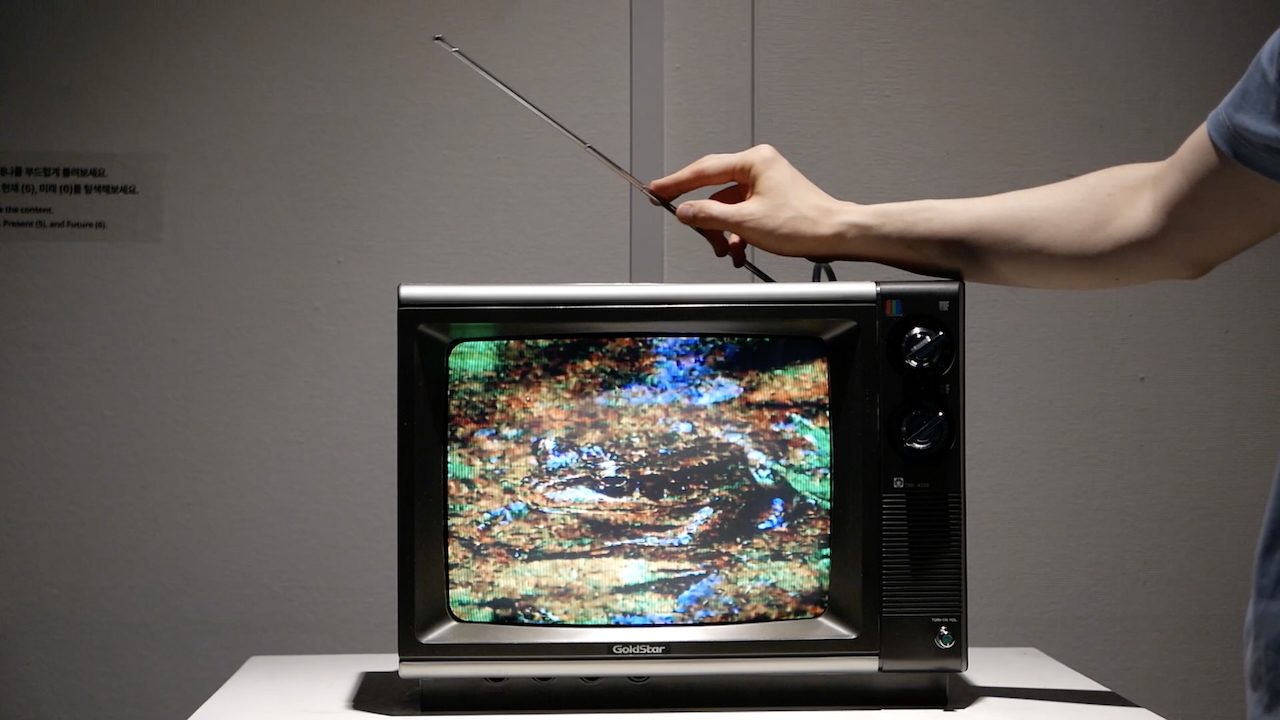}
  \caption{Diffusion TV installation}
  \Description{Image of the Diffusion TV installation where a person is interacting with the work holding the antenna of a CRT TV placed on a pedestal.}
  \label{fig:installation}
\end{figure}

As shown in Figure \ref{fig:installation}, Diffusion TV was first exhibited at the International Symposium on Electronic/Emerging Art 2025, held from May 23 to 29, 2025, at the Hangaram Design Museum in Seoul, South Korea, where it was open to the public.

The following reflections are based on the author’s exhibition observations and video recordings rather than a structured empirical evaluation. Audiences typically learned the interaction through exploration, with prior familiarity with CRT TVs shaping engagement: older participants navigated intuitively, while younger viewers sometimes struggled with the interface. The antenna-based denoising interaction was generally perceived as intuitive and engaging; however, some participants did not fully explore sequential content or expected additional system behaviors. Overall, the tuning metaphor supported an experiential engagement with the generative process and encouraged thematic interpretation without explicit instruction.

Video documentation of the Diffusion TV installation and audience interactions is available at \url{https://sihwapark.com/Diffusion-TV}.

\section{Conclusion and Future Work}
Diffusion TV demonstrates the potential of embodied, metaphorical engagement as an alternative mode of XAI that moves beyond explicit technical explanation in conventional 2D GUI-based approaches. While the tangible interaction with the denoising process supports intuitive and experiential engagement, it does not guarantee precise comprehension of diffusion models, and audience responses varied depending on prior familiarity with AI. Observations were based on informal feedback and qualitative reflection rather than systematic evaluation. Future work will involve structured audience studies to examine how embodied interaction shapes public interpretation of generative AI systems. In addition, the system will be extended toward real-time or hybrid inference pipelines to maintain responsive interaction while enabling more flexible generative behavior, alongside further refinement of tangible controls to support richer exploration of the generative process.
\begin{acks}
This project was undertaken thanks in part to funding from the Connected Minds Program, supported by Canada First Research Excellence Fund, Grant \#CFREF-2022-00010.
\end{acks}

\bibliographystyle{ACM-Reference-Format}
\bibliography{ref}

\end{document}